# Evidence for Defect-induced Superconductivity up to 49 K in $(Ca_{1-x}R_x)Fe_2As_2$

L. Z. Deng[1], B. Lv[1,*], K. Zhao[1], F. Y. Wei[1], Y. Y. Xue[1], Z. Wu[1] and C. W. Chu[1,2,†]

1. Texas Center for Superconductivity and Department of Physics, University of Houston, Houston, Texas 77204-5002, USA.

2. Lawrence Berkeley National Laboratory, Berkeley, California 94720, USA.

*Present address: Department of Physics, University of Texas at Dallas, Richardson, Texas 75080, USA.

†Corresponding author: cwchu@uh.edu

## Abstract

To explore the origin of the unusual non-bulk superconductivity with a $T_c$ up to 49 K reported in the rare-earth-doped $CaFe_2As_2$, the chemical composition, magnetization, specific heat, resistivity, and annealing effect are systematically investigated on nominal $(Ca_{1-x}R_x)Fe_2As_2$ single crystals with different x's and R = La, Ce, Pr, and Nd. All display a doping-independent $T_c$ once superconductivity is induced, a doping-dependent low field superconducting volume fraction $f$, and a large magnetic anisotropy η in the superconducting state, suggesting a rather inhomogeneous superconducting state in an otherwise microscale-homogenous superconductor. The wavelength dispersive spectroscopy and specific heat show the presence of defects which are closely related to f, regardless of the R involved. The magnetism further reveals that the defects are mainly superparamagnetic clusters for R = Ce, Pr, and Nd with strong intercluster interactions, implying that defects are locally self-organized. Annealing at 500 °C, without varying the doping level x, suppresses f profoundly but not the $T_c$. The above observations provide evidence for the crucial role of defects in the occurrence of the unusually high $T_c$ ~ 49 K in $(Ca_{1-x}R_x)Fe_2As_2$ and are consistent with the interface-enhanced superconductivity recently proposed.

## Introduction

The Fe-based layered pnictides and chalcogenides upon doping or under pressure constitute an interesting superconductor class with a transition temperature $T_c$ as high as 57 K [1], second only to that of ≤ 134 K at ambient [2] or 164 K under pressure [3] of the layered cuprate class. These superconductors comprise four families with their respective maximum $T_c$s at ambient as: 1111 (RFeAsO)—57 K [1]; 122 ($AEFe_2As_2$ or AE122)—38 K [4,5] ; 111 (AFeAs)—18 K [6]; and 11 (FeSe)—10 K [7], where R = rare-earth, AE = alkaline earth, and A = alkaline, respectively. While doping by K or Na in the AE-sites generates bulk superconductivity with a maximum $T_c$ ~ 33–38 K as evidenced calorimetrically in all AE122 [8], pressure induces superconductivity with a similar maximum $T_c$ only in Ba122 and Sr122 [9]. Ca122 appears to behave differently from other 122 members. It becomes superconducting under quasi-hydrostatic pressure, but only below 12 K [10] and not under hydrostatic pressure [11]. It also exhibits a complex phase diagram with an unusual first-order tetragonal collapsed phase transition below ~ 160 K in the presence of pressure or on doping [12].

Single crystals of $(Ca_{1-x}R_x)Fe_2As_2$ [(Ca,R)122] with the 122 layer structure were found to show superconductivity with an unexpectedly high onset temperature $T_c$ up to 49 K when Ca is partially replaced by rare-earth elements R = La, Ce, Pr, and Nd of comparable ionic radii up to their respective solubility limits. Such a $T_c$ is the highest among the bulk Fe-pnictides and Fe-chalcogenides with the same structures at ambient or under pressure [13-16]. The superconductivity in these single crystals were later found to be non-bulk and highly anisotropic, e.g. the low field shielding volume fraction $f \equiv 4\pi\chi_{ZFC}^{ab}$, where ZFC stands for Zero-Field-Cool, is less than 10% and the magnetic anisotropy is up to 200 at 5 K for R = Pr. In spite of the apparent homogeneity of these single crystals within our wavelength dispersive spectroscopy (WDS) resolution of 1 μm, the superconductivity is clearly inhomogeneous. Three possible causes have subsequently been advanced to account for such unusual heterogeneous superconductivity. They are: (1) minute inclusion of the superconducting phase, such as RFeAsO$_{1-\delta}$ with a T$_c$ up to 57 K [1,17]; (2) doping inhomogeneity, *i.e.* only a small fraction of the crystals has the required doping level x=R/(Ca+R)[13,18]; or (3) superconductivity associated with defects present in the single crystals [17].

To differentiate among the above possibilities, the stoichiometry, structure, magnetization, resistivity, and specific heat of more than twenty (Ca,R)122 single crystals slightly doped with both the magnetic and non-magnetic rare-earth elements undergoing different annealing conditions were investigated. We found that all single crystals examined display noticeable deviations from the 122-stoichiometry. These defects are also detected in our magnetization and specific heat measurements. For R = Ce, Pr, and Nd, the magnetization data further show superparamagnetic clusters with strong interactions among the defects, suggesting the local ordering of the defects. When the defects are varied by annealing without perturbing the carrier density, the superconducting volume fraction is observed to change drastically but not the T$_c$. The above results, together with the very large 2D-like anisotropy detected in the superconducting state of all samples, suggest that the enhanced T$_c$ may be interfacial in nature.

**Experimental**

The single crystals of (Ca,R)Fe$_2$As$_2$, where R = La, Ce, Pr, and Nd, were successfully grown from self-flux. The FeAs precursor was first synthesized from stoichiometric amounts of Fe (99.999 + % from Aldrich) and As (99.9999% from Alfa) inside the silica tube at 800 °C for 30 h. Then R-pieces (99.9% from Alfa) and Ca-pieces (99.99% from Alfa) were mixed with FeAs according to the ratio of (R+Ca)/FeAs = 1/4 and placed in an alumina crucible inside a silica tube sealed under reduced Ar atmosphere. The silica tube was subsequently sealed inside a larger silica tube under vacuum to prevent the sample from getting into contact with air if the first tube failed. The assembly was then put inside a box furnace, heated to 1,200 °C for 8 h, and then cooled to 980 °C slowly at 2 °C/hr. The sample was finally furnace-cooled to room temperature by turning off the power. Single crystals with the flat shiny surface up to 5 mm × 5 mm size were easily cleaved from the melt. All of the preparative manipulations were carried out in a purified argon atmosphere glove box with a total O$_2$ and H$_2$O level <1 ppm. The systematic annealing was carried out on (Ca,R)122 crystals by sealing the crystal in an evacuated quartz tube, and heating it in a furnace at 500 °C under vacuum for a certain period of time before quenching in ice water. After this, the sample was characterized and then carefully cleaned for further annealing, with the cumulative annealing time, t, up to 100 hours. The X-ray diffraction patterns of the as-synthesized samples were obtained using a Rigaku DMAX III-B diffractometer. The chemical analyses were performed using WDS on a JEOL JXA-8600 electron microprobe analyzer with 1 μm spot size giving an estimated systematic deviation below 0.5%. The magnetic measurements were carried out employing the 5 T Quantum Design Magnetic Property Measurement System (MPMS). The four-lead resistivity and specific heat measurements were performed in the 7 T Quantum Design Physical Property Measurement System (PPMS) with a Helium3 option at temperatures down to 0.4 K.

## Results and Discussion

*1. Uniformity and Stoichiometry*

X-ray diffractions were used to characterize the samples and showed no second phase within its resolution. WDS has been employed to examine the chemical uniformity and the stoichiometry of the samples. The presence of lattice defects in these samples is evident. For the convenience of discussion, the results are represented by the formulas of $Ca_{1-x}Pr_xFe_{2+y}As_{2-z}$ for $0 \leqslant x \leqslant 0.125$ and $Ca_{1-x-a}La_xFe_{2+y}As_{2-z}$ for $0 \leqslant x \leqslant 0.185$ are shown in the ternary plots in Figs. 1a and 1c, respectively. Off-stoichiometry in these sample from the ideal 122, i.e. non-zero y or z, is clearly evident. The relative deviation y/2 and z/2 are at the level of < 10% for (Ca,Pr) and < 5% for (Ca,La)122 as shown in Tables I and II. The random doping-spread is expressed as $\Delta x_i = x_i - x_{avg}$ in Fig. 1b, where $x_i$ and $x_{avg}$ are the WDS value at the $i_{th}$ point and its average over the same crystal, respectively. The histogram in Fig. 1b demonstrates that the macroscopic x-spread is on the order of 0.002 and no deviation larger than ±0.01 has been detected. While nano-scale Pr non-uniformity may still be possible in principle within the limited spatial-resolution of WDS, ≈ 1 μm, the fast ion-diffusion expected during synthesis make this case unlikely. The absence of Pr cluster is supported experimentally by the scanning tunneling microscopy (STM) results [19]. Therefore, doping inhomogeneity can hardly cause the $f \ll 1$ observed.

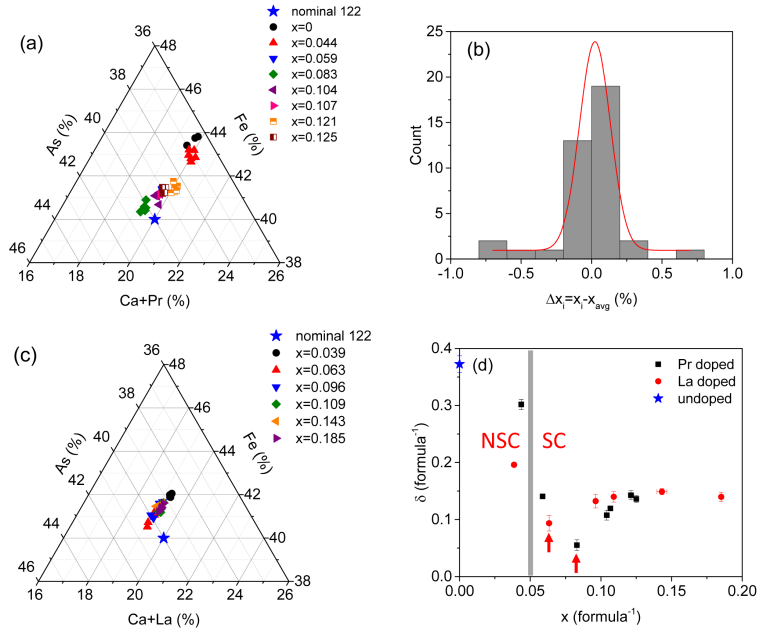

Fig. 1. (a) Ternary plot of (Ca+Pr) vs. Fe vs. As for $Ca_{1-x}Pr_xFe_{2+y}As_{2-z}$; (b) The spread of the local doping levels from WDS measurements against the average over the same crystal. The red line represents a Gaussian fitting for the data; (c) Ternary plot of (Ca+La) vs. Fe vs. As for $Ca_{1-x-a}La_xFe_{2+y}As_{2-z}$; (d) Stoichiometry distance vs. doping for Pr- and La-doped samples, and undoped sample. Vertical arrows indicate the minimum points of δ for both Pr and La doping. The gray bar represents the boundary between the superconducting region and the non-superconducting region.

| Pr | (2+y) | Δ(2+y) | $\frac{y}{2}$ | (2-z) | Δ(2-z) | $\frac{z}{2}$ |
|---|---|---|---|---|---|---|
| 0 | 2.157 | 0.008 | 7.85% | 1.810 | 0.028 | 9.50% |
| 0.044 | 2.124 | 0.012 | 6.20% | 1.833 | 0.006 | 8.35% |
| 0.059 | 2.070 | 0.004 | 3.50% | 1.948 | 0.004 | 2.60% |
| 0.083 | 2.028 | 0.011 | 1.40% | 1.999 | 0.009 | 0.05% |
| 0.104 | 2.054 | 0.010 | 2.70% | 1.965 | 0.007 | 1.75% |
| 0.107 | 2.060 | 0.003 | 3.00% | 1.956 | 0.004 | 2.20% |
| 0.121 | 2.071 | 0.010 | 3.55% | 1.925 | 0.007 | 3.75% |
| 0.125 | 2.068 | 0.007 | 3.40% | 1.948 | 0.005 | 2.60% |

Table I. (2+y) and (2-z) range for (Ca,Pr)122 single crystals.

| La | (2+y) | Δ(2+y) | $\frac{y}{2}$ | (2-z) | Δ(2-z) | $\frac{z}{2}$ |
|---|---|---|---|---|---|---|
| 0.039 | 2.098 | 0.004 | 4.90% | 1.937 | 0.003 | 3.15% |
| 0.063 | 2.047 | 0.015 | 2.35% | 1.999 | 0.016 | 0.05% |
| 0.096 | 2.066 | 0.015 | 3.30% | 1.979 | 0.016 | 1.05% |
| 0.109 | 2.070 | 0.010 | 3.50% | 1.970 | 0.008 | 1.50% |
| 0.143 | 2.075 | 0.005 | 3.75% | 1.970 | 0.008 | 1.50% |
| 0.185 | 2.070 | 0.008 | 1.50% | 1.970 | 0.008 | 1.50% |

Table II. (2+y) and (2-z) range for (Ca,La)122 single crystals.

A doping-dependent off-stoichiometry is also clearly observed in Figs. 1a and 1c: all data points cluster around the (Ca + Pr)/(Ca+Pr+Fe+As) ~ 19-20% line with z ≈ y > 0; and around the (Ca + La)/(Ca+La+Fe+As) ~ 19% line with z ≈ y > 0, although Pr is magnetic and La is not. They all display an As-deficiency and a Fe-excess, as represented by $Ca_{1-x}Pr_xFe_{2+y}As_{2-z}$ and $Ca_{1-x-a}La_xFe_{2+y}As_{2-z}$, where y and z are of comparable values. The data suggest that the defects are anti-sites. The overall defect density associated with the non-zero a, y, and z can be represented by δ=$a$+y+z. Two interesting features emerge and are shown in Fig. 1d: (1) δs are comparable for both (Ca,Pr)122 and (Ca,La)122 and (2) a minimum of δ appears near the metal and superconductor boundary in both cases. This non-monotonic x-dependence suggests that the physics associated with the defects as well as their microstructure are likely to be different in these two regions.

In the superconducting region, two superconducting transitions are easily detected resistively at $T_{c1}$ ~ 20 K and $T_{c2}$ ~ 50 K, respectively, as exemplified in Figs. 2a and 2b for (Ca,Pr)122 and (Ca,La)122. High pressure effect on the two resistive transitions was investigated by W. Uhoya et al. [20] and S. R. Saha et al. [21]. This two-step transition is also observed in magnetization measurements [13,17]. Both $T_{c1}$ and $T_{c2}$ are found to be not sensitive to the doping x (Figs. 2c and 2d), in strong contrast to the superconducting volume fraction f, which increases rapidly with x in the superconducting region (Fig. 3). These demonstrate that the two superconducting transitions are not caused by the carrier density change due to doping but possibly by the local nanostructure change due to defects.

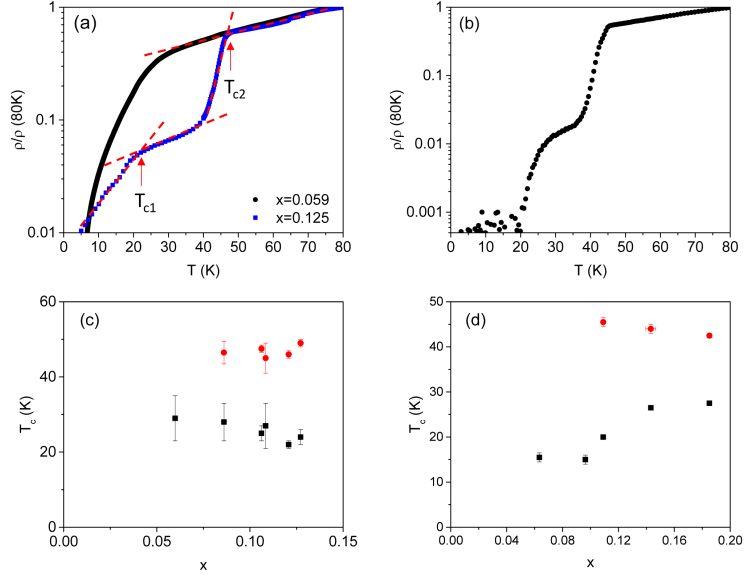

Fig. 2. ρ/ρ (80K) for (a) (Ca,Pr)122 with x=0.059 and 0.125; (b) (Ca,La)122 with x=0.143. The dashed lines and arrows show the way $T_c$ was decided. $T_c$ as a function of doping for (c) (Ca,Pr)122 samples and (d) (Ca,La)122 samples. Black squares: $T_{c1}$; red circles: $T_{c2}$.

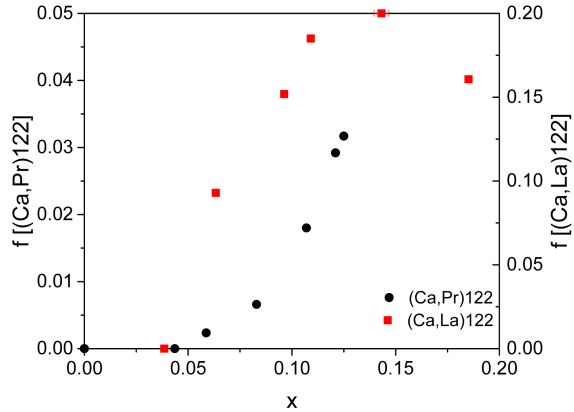

Fig. 3. Superconducting volume fraction as a function of doping for (Ca,Pr)122 and (Ca,La)122 samples at 5 K.

## 2. Magnetization

Doping, in general, modifies the compound properties via carrier density change and/or microstructure change. Some of these changes can be reflected in their magnetic properties. Clearly, the magnetic properties of the single crystalline samples investigated depend on R. For instance, the magnetic moments (M) of the (Ca,R)122 for R = Ce, Pr, and Nd are about ten times that for R = La at room temperature. In spite of the difference, different Rs induce superconductivity with the similar $T_c$ of ~ 45 (La) – 50 K (Pr), as exhibited in Figs. 2c and 2d.

While M of samples with R-doping (R=Ce, Pr, and Nd) shows a temperature-dependent paramagnetic behavior, that of La-doped sample is almost temperature independent above ~ 30 K as expected (Fig. 4). However, the overall M(T) of the (Ca,Pr)122 samples is not consistent with that of free $Pr^{+3}$ as shown in Fig. 4. We have therefore investigated the M(T,H) of (Ca,Pr)122 and (Ca,La)122 infested with defects as discussed earlier.

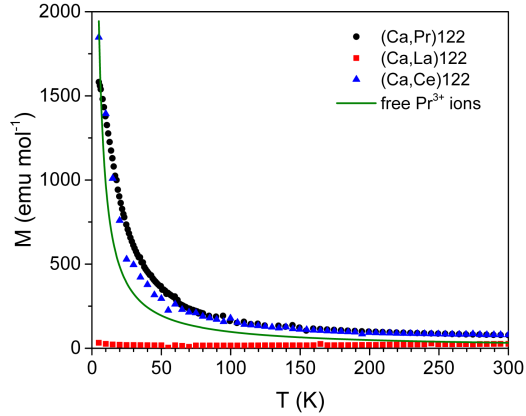

Fig. 4. M vs. T under H = 5 T for Pr (x=0.121), La (x=0.185), and Ce (x=0.182) doped Ca122 samples and estimated contribution of the free $Pr^{3+}$ ions.

Magnetic ordering is avoided in the iron-based superconductors through a delicate balance among various Fe 3d sub-bands. Net moments are expected surrounding lattice defects, where the broken Fe-As bonds disturb the balance. Magnetic clusters around lattice vacancies were theoretically proposed [22,23]. Various bulk cluster-orderings have been consequentially proposed/observed in A-Fe-Se system [24-26]. We conjecture that the defects in (Ca,R)122 may appear as superparamagnetic clusters, which can be identified by examining the isothermal M-H loops or the Neel relaxation. They probe the cluster volume V through the competitions between the magnetic energy mH and the thermal energy $k_B T$, where m is the magnetic moment for each cluster and $k_B$ is the Boltzmann constant. No hysteresis has been detected in the M(H,T) of (Ca,Pr)122 up to 5 T down to 5 K within our resolution. The M can thus be considered as an equilibrium property of the sample and be described in terms of the Langevin function [27], i.e. M = nm[1/tanh(p)-1/p] with p =mH/$k_B$T, where n is the density of superparamagnetic clusters and m the magnetic moment for each cluster. The Langevin function fits roughly the data as shown by the dashed curves in Fig. 5. However, the M-values for the sample of (Ca,Pr)122 with x = 0.125 so-calculated are lower than the experimental results at low temperatures but become greater at higher temperature, suggesting a ferromagnetic-like inter-cluster interaction. We have therefore included the contributions from the paramagnetic $Pr^{+3}$-ions and the possible magnetic interaction. The resulting M becomes M = nm[1/tanh(q)-1/q]+ H·$\chi_{Pr}$, where H·$\chi_{Pr}$ = x·H $\frac{N(3.5\mu_B)^2}{3k_B(T+T_0)}$ is the contribution from the paramagnetic $Pr^{+3}$ with a moment of 3.5$\mu_B$ and q=mH/$k_B(T+T_0)$, where $T_0$ is the effective Curie-Weiss temperature for the superparamagnetic clusters. This M fits the data well as the solid red-curves shown in Fig. 5 for (Ca,Pr)122 with x =0.125 well into the superconducting region. The extracted values for n, m, and $T_0$ are shown in Figs. 6 and 7 for later discussions. Similar M(H)-isotherms are obtained for the (Ca,Ce)122 and (Ca, Nd)122 samples (Table III), although the n extracted from the (Ca,La)122 data shows a cluster density 100 times lower.

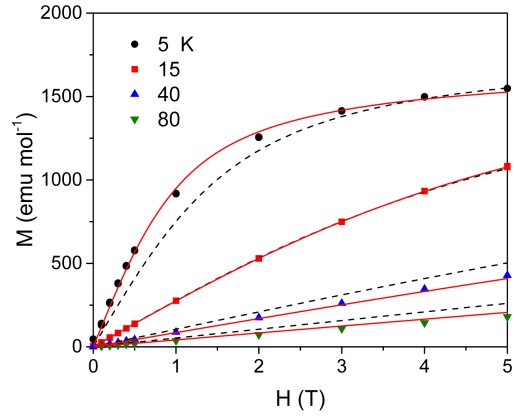

Fig. 5. M-H loops for (Ca,Pr)122 with x = 0.125. Black dashed lines represent fittings by Langevin function and red solid lines represent fittings by modified Langevin function.

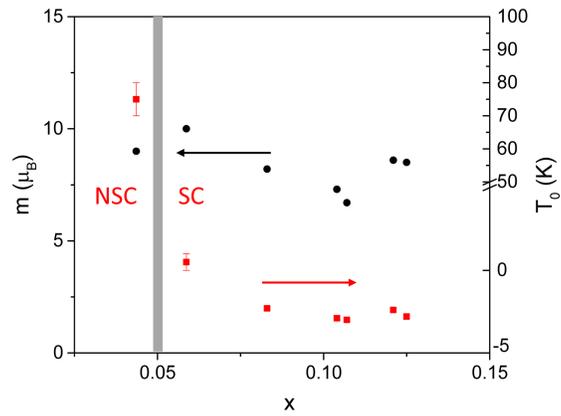

Fig. 6. m (black circles) and $T_0$ (red squares) as a function of doping for (Ca,Pr)122. The gray bar represents the boundary between the superconducting region and the non-superconducting region.

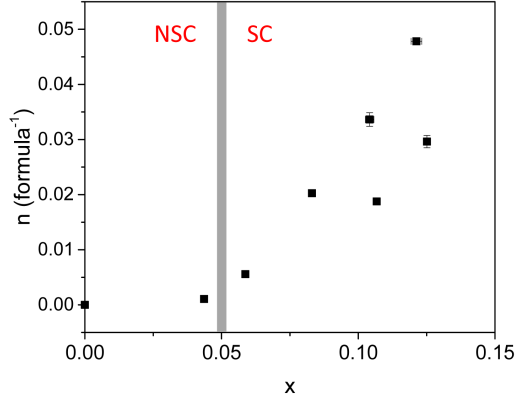

Fig. 7. Superparamagnetic cluster density as a function of doping for (Ca,Pr)122 samples. The gray bar represents the boundary between the superconducting region and the non-superconducting region.

| R | Pr | | | | Ce | Nd | La | | | |
|---|---|---|---|---|---|---|---|---|---|---|
| Doping | 0.044 | 0.059 | 0.104 | | 0.182 | 0.064 | 0.039 | 0.097 | 0.143 | 0.185 |
| Exp | MH | MH | MH | $C_p$ | MH | MH | | $C_p$ | | |
| n(Def) | $1.07 \times 10^{-3}$ | $5.6 \times 10^{-3}$ | $2.48 \times 10^{-2}$ | $2.45 \times 10^{-2}$ | $3.20 \times 10^{-2}$ | $2.60 \times 10^{-2}$ | $1.6 \times 10^{-5}$ | $4 \times 10^{-5}$ | $8.9 \times 10^{-4}$ | $7.5 \times 10^{-4}$ |
| Stddev | $1.4 \times 10^{-4}$ | $1 \times 10^{-4}$ | $4 \times 10^{-4}$ | $4 \times 10^{-4}$ | $2 \times 10^{-4}$ | $3 \times 10^{-4}$ | $3 \times 10^{-6}$ | $3 \times 10^{-6}$ | $5 \times 10^{-5}$ | $3 \times 10^{-5}$ |

Table III. Summary of defect density (/formula) for (Ca,R)122 samples. "Exp" indicates experimental method, "n(Def)" stands for defect density, and "Stddev" is standard deviation. "MH" indicates that the defect density is calculated from magnetization measurement data and "$C_p$" indicates that the defect density is deduced from heat capacity results.

It is interesting to point out that the moment per cluster m in the (Ca,Pr)122 crystals tested at different temperatures is m = 8±2 $\mu_B$, as shown in Fig. 6, lending validity to our analysis. The total number of ions in a cluster is on the order of 10-100, well below the commonly accepted number for ferromagnetic domains [28]. In other words, the data confirm that they are superparamagnetic clusters, but very unlikely associated with nanoscale impurities. This confirms the above conjecture that the magnetic background of (Ca,Pr)122 is dominated by lattice defects in the FeAs layers.

From Fig. 6, the non-zero $T_0$ clearly shows that significant interaction exists between the magnetic clusters. Fig. 7 shows that n is practically zero for all non-superconducting as-synthesized (Ca,Pr)122 crystals for x < 0.05, despite their rather large non-stoichiometry observed. For x > 0.05 in the superconducting region, n increases with x, similar to what we have observed for the superconducting volume fraction f of the as-synthesized samples (Fig. 3). The observation suggests that defects and superconductivity in these crystals are closely related and the mesostructures surrounding the lattice defects in the non-superconducting region are different from those in the superconducting region.

3. *Specific heat*

The magnetization of (Ca,R)122 for magnetic R has been found to be about 100 times greater than that for the nonmagnetic R = La; and superparamagnetic cluster formation associated with lattice defects has been detected in the former but not the latter. At the same time, a similar $T_c$ occurs in all the samples. If lattice defects are the driving force for superconductivity, they should also exist in (Ca,La)122. An experimental method that does not involve magnetism is needed. We have therefore decided to employ the calorimetric technique that can detect the effects of both magnetic and non-magnetic defects on the electronic and phonon energy spectra simultaneously.

The specific heat $C_p$ has been measured for several (Ca,R)122 crystals at various doping levels, both before and after annealing. In all cases, the $C_p/T$ observed fit $\gamma_0 + \beta \cdot T^2$ well below 20 K except for the Schottky anomaly. As usual, the slope $\beta$ and the zero-temperature interception $\gamma_0$ of the $C_P/T$ - $T^2$ plot will provide information about the electronic and phonon characteristics of the samples, respectively. We have also developed a method to extract information about the defect density n from the Schottky anomaly as described later.

The variation of $\gamma_0$ with x for the as-synthesized crystals with R = La is shown in Fig. 8a. A linear dependence of $\gamma_0$ on x is evident with a slope $d\gamma_0/dx \approx 0.2$ J/mol·K$^2$ throughout the superconducting region, i.e. 0.06 < x < 0.21. This suggests that R dopes into the (Ca,La)122 single crystals continuously and changes their electron energy spectra, while leaving $T_{c1}$ and $T_{c2}$ almost constant (Fig. 2d). The observations also show that the $T_{c1}$ and $T_{c2}$ cannot be a direct result of doping. On the other hand, the doping effect of x on $\beta$ is very different. As shown in Fig. 8b, $\beta$ exhibits a ~40% jump abruptly as the $T_{c2}$-transition appears near x~0.1 and varies very little for x> 0.1, *i.e.* less than a few percent over $0.109 \leq x \leq 0.185$ in the superconducting region of $T_{c2}$, similar to the x-insensitive $T_{c1}$ and $T_{c2}$ (Figs. 2c and 2d) and lattice defects due to off-stoichiometry $\delta$ (Fig. 1d). This further demonstrates that $\beta$ is closely related to the defects in the samples through their effects on the phonons.

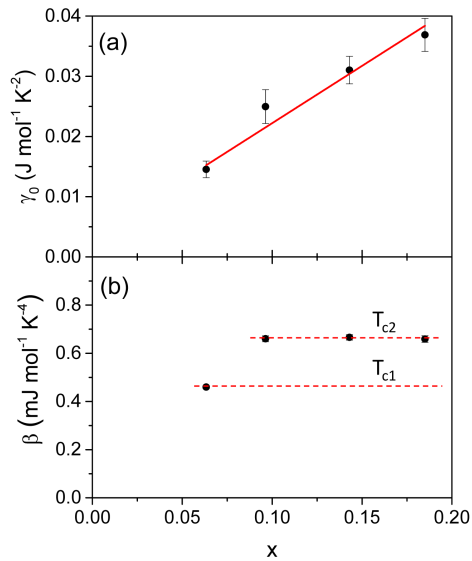

Fig. 8. (a) $\gamma_0$ vs. doping for (Ca,La)122 single crystals; (b) $\beta = d(C_p/T)/dT^2$ vs. doping. The red solid line represents the linear fitting for the data and the red dashed lines represent different doping regions for crystals with different $T_c$s.

Concurrently with our work, S. R. Saha *et al.* observed a reduction in the diamagnetic signal for 700 °C annealed (Ca,La)122 samples [14] and Okada *et al.* reported the decrease of superconducting volume fraction for 400 °C annealed (Ca,Pr)122 samples [29]. We have adopted 500 °C annealed (Ca,La)122 with x = 0.185 for different periods of time to control the defect density and examine its effect on f(2 K), $\beta$, and $\gamma_0$. We found that annealing has little effect on $\gamma_0$ (Fig. 9a) while suppressing f and $\beta$ rapidly (Figs. 9a and 9b) due to the removal of defects. The former shows that 500 °C annealing does not alter the electronic structure or doping of the sample as expected. The latter confirms our conjecture that defects play a crucial role in the superconductivity with high $T_c$ in (Ca,La)122 single crystals. The clear monotonic relationship between $\beta$ and the superconducting volume fraction f for (Ca,Pr)122 and (Ca,La)122 shown in Fig. 10 further support the strong correlation between the defects and high $T_c$ in the (Ca,R)122 system.

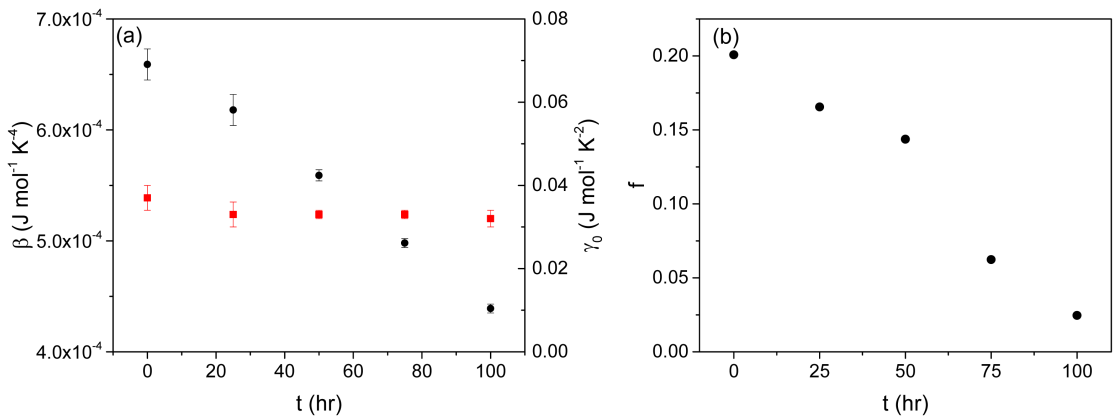

Fig. 9. (a) $\beta$ (black circles) and $\gamma_0$ (red squares) under different annealing time at 500 °C; (b) Volume fraction f(2K) vs. annealing time at 500 °C for (Ca,La)122 (x = 0.185) sample.

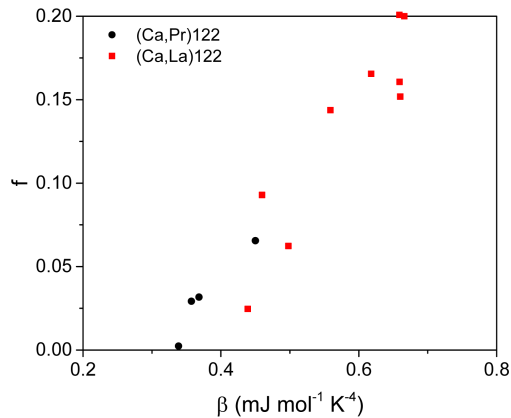

Fig. 10. f vs. $\beta$ for as-synthesized (Ca,Pr)122 samples and (Ca,La)122 samples with different doping and under different annealing time at 500 °C.

As described earlier, n can be determined magnetically through the superparamagnetic measurements for R = Ce, Pr, and Nd using the Langevin formula but this is difficult for R = La. We have therefore developed a procedure to determine n calorimetrically by analyzing the Schottky anomaly appearing in $C_p$ by employing the multi-level Schottky formula,

$$\frac{CH}{T} = \frac{nk_B H}{T}\left[\frac{s^2 e^s}{(e^s-1)^2} - (2J+1)^2 \frac{s^2 e^{(2J+1)s}}{(e^{(2J+1)s}-1)^2}\right], s = \frac{g\mu_B H}{k_B T}$$

, where n is the number of Schottky centers corresponding to the defect density in our case, g is the effective g factor for the defect, and J is the spin [30]. To test the procedure, we have determined n of (Ca,Pr)122 for x = 0.104 both magnetically and calorimetrically as displayed in Fig. 11. The values of *n* so-deduced are in good agreement with those determine magnetically (Table III), demonstrating the validity of the procedure adopted. The values of n for (Ca,La)122 are then deduced calorimetrically and listed in Table III.

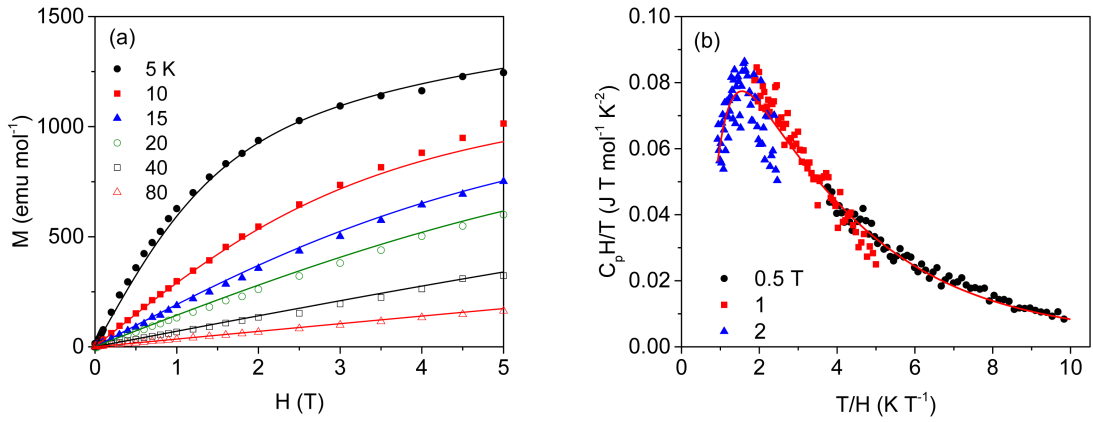

Fig. 11. (a) M-H for (Ca,Pr)122 (x=0.104); (b) $C_p H/T$ vs. T/H for (Ca,Pr)122 (x=0.104) at H up to 2T.

The n of (Ca,La)122 determined calorimetrically shows the expected drop with annealing due to the associated reduction in defects (Fig. 12). Similar to (Ca,Pr)122, the n of (Ca,La)122 so-determined reveals a clear difference between the superconducting and non-superconducting samples, e.g. > $10^{-4}$ for the former and < $10^{-4}$ for the latter. The n-f correlation, i.e. f increases with n (Fig. 13), is similar to (Ca,Pr)122. The observation clearly demonstrates that the superconductivity with a high $T_c$ is closely coupled with the defects. In spite of the above similarities between (Ca,Pr)122 and (Ca,La)122, the n of the superconducting (Ca,Pr)122 is about ~ 40 times that of (Ca,La)122 with similar maximum $T_c$. The difference in f can be caused by the difference in ionic sizes of Pr and La. However, the difference alters the $T_c$ of the two only negligibly, suggesting similar defect-induced microstructures in the superconducting samples for all Rs.

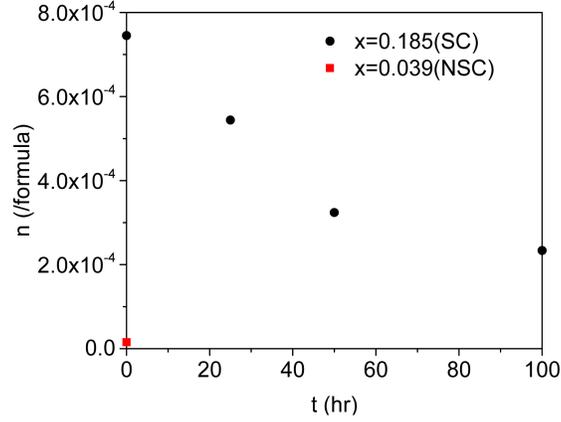

Fig. 12. Defect density vs. annealing time at 500 °C for SC and NSC (Ca,La)122 samples.

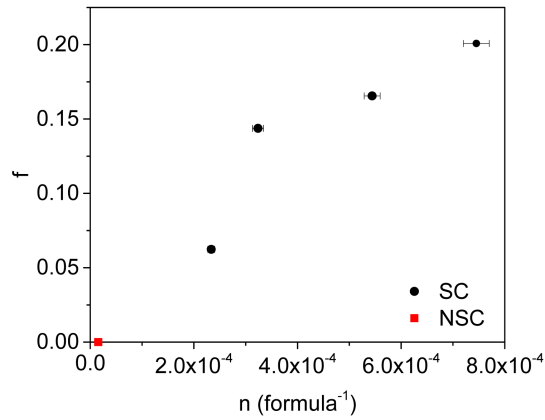

Fig. 13. Superconducting volume fraction (2K) vs. defect density for SC (x=0.185) and NSC (Ca,La)122 (x=0.039) samples.

4. *Defect-induced superconductivity*

The above results reveal clearly a close relationship between the superconductivity in (Ca,R)122 with an unusually high $T_c$ and the defects in them, whether these defects are caused by off-stoichiometry, chemical doping, and/or off-thermal equilibrium during formation, supporting our conjecture that the superconductivity detected is induced by defects once the threshold is reached. The f observed scales well with the defect concentration n. However, direct evidence to demonstrate that the enhanced $T_c$ is caused by the interfaces between different defect-phases, as previously suggested by us [17,31,32] is yet to be provided.

It is known that the magnetic exchange-field between the defects can have significant value only if the defects are spatially ordered. We found that $T_0$ of the as-synthesized (Ca,Pr)122 changes systematically with x from ~ - (2 - 3) K in the superconducting region for x > 0.06 to ~ + 1 K at x = 0.059 to ~ + 75 K at x = 0.044, where the superconducting shielding is suppressed to below the noise floor. The raw low-field susceptibility $\chi_0 = dM/dH |_{H\to 0}$ is plotted as the function of 1/T in Fig. 14. While the $\chi_0$ of Crystal x =

0.059 appears as a straight line, which is expected for non-interactive clusters, the bump around 15 K of Crystal x=0.044 demonstrates a well developed AFM-order. Also, while the data of Crystal x=0.125 seem to be rather close to that of Crystal x=0.059, the systemetic deviations at 5 K and 10 K are much larger than the experimental uncertainties. Noticeable ferromagnetic interactions are unavoidable. The effects of the annealing on $T_0$ are much weaker and show significant data fluctuations. One of the annealed crystals does show a slightly positive $T_0 \approx 2$ K. Such a noticeable $T_0$ and the associated intercluster fields require cluster ordering in the mean-field-approximation.

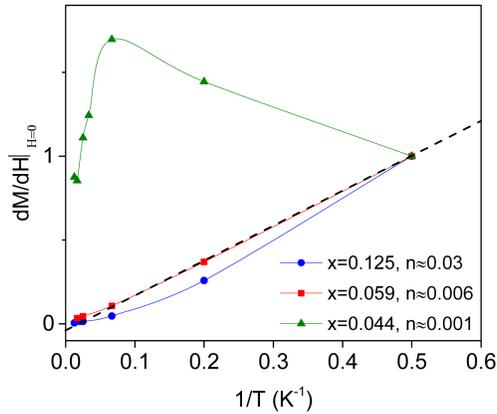

Fig. 14. The initial *dc* susceptibility $dM/dH|_{H=0}$ over $2\,K \leq T \leq 50\,K$ for several as-synthesized (Ca,Pr)122 crystals. The black dashed line corresponds to the $c \propto 1/T$ of non-interacting clusters.

To further explore the issue, the anisotropy of the superconductive screening is measured. The demagnetizing enhancement associated with the anisotropy is a direct measurement of the topology of the corresponding superconductive parts. In the case of the low *f*, superconductive domains are well separated, and more-or-less follow the geometry of the mesostructure of the defects. For isolated point defects, the superconductivity should be isotropic. The screening should show the rod-like anisotropy, *e.g.* $|\chi_{H//a}| \approx |\chi_{H//b}| \gg |\chi_{H//c}|$, with the rod axis lying along the *c* direction; or the disk-like anisotropies, *e.g.* $|\chi_{H//c}| \gg |\chi_{H//a}| \approx |\chi_{H//b}|$, with the disk-like mesostructure located on the *ab* plane. It has been reported that the low-field screening of (Ca,Pr)122 is extremely anisotropic with the ratio $\chi_{H//c}/\chi_{H//ab}$ on the order of 20-100, but no in-plane anoisotropy is noticeable [17]. This further supports the above judgment that point defects, rather than nanoscale impurities, are the main lattice defects in (Ca,R)122 single crystals.

The most reasonable interpretation for the above observation will be the ordered defects, especially interfacial superconductivity. This situation is rather similar to that observed in K-Fe-Se [33], where ordered defects may form different mesoscopic phases. It has been further suggested that the superconductivity is enhanced by the interfaces between such different phases, though it is still an open question. With the unusually high $T_c \approx 49$ K in (Ca,R)122, which is higher than that in all other reported compounds of Ca-R-Fe-As, the data presented above are in line with the interface superconductivity model [17,31,32]. To verify that this is a universal property for all (Ca,R)122, several (Ca,Pr)122 and (Ca, La)122 crystals were tested as exemplified in Fig. 15. The screening anisotropy all exceed ~ 7 due to the demagnetization factor of the sample, in support of the interfacial enhanced $T_c$ scenario. This can be understood only if the superconductivity occurs in the self-organized defects.

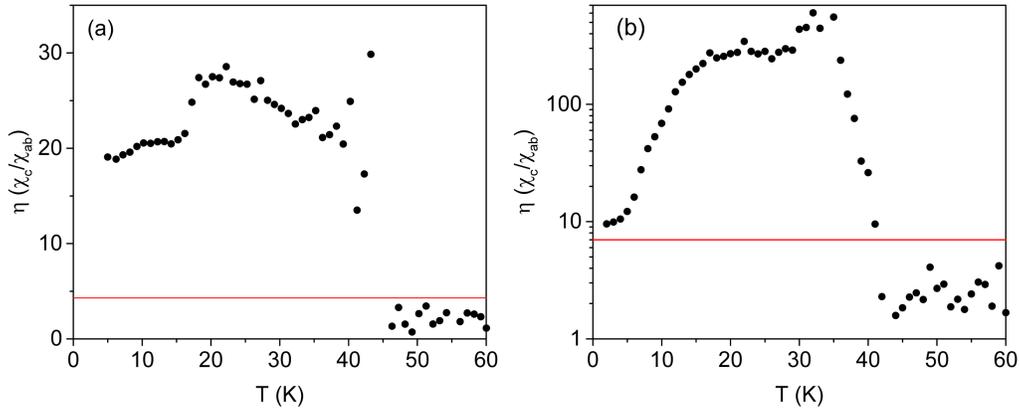

Fig. 15. Magnetic anisotropy measurements for (a) (Ca,Pr)122 and (b) (Ca,La)122. Red lines represent the sample geometric anisotropy.

## Summary

In conclusion, we have measured and analyzed the chemical stoichiometry, magnetization, resistivity, and specific heat of the as-synthesized and 500 °C annealed (Ca,R)122 single crystalline samples. We found that doping alters the defect density and the superconducting volume fraction drastically but not the $T_c$ for as-synthesized samples. The same effects were also observed for 500 °C annealing, without changing the chemical composition, demonstrating the close relationship between the defect density and the superconducting volume fraction. Together with the large magnetic anisotropy and the ordered nature of defects, the above results suggest that the superconductivity with an enhanced $T_c$ is induced by interfaces, although details of the interfaces are yet to be determined.

## Acknowledgments

The work in Houston, Texas, is supported, in part, by U.S. Air Force Office of Scientific Research Grants FA9550-09-1-0656 and FA9550-15-1-0236, the T.L.L. Temple Foundation, the John J. and Rebecca Moores Endowment, and the State of Texas through the Texas Center for Superconductivity at the University of Houston.

## References

[1] Z. Wei, H. Li, W. Hong, Z. Lv, H. Wu, X. Guo, and K. Ruan, Journ. of Supercond. and Nov. Mag. **21**, 213 (2008).
[2] A. Schilling, M. Cantoni, J. D. Guo, and H. R. Ott, Nature **363**, 56 (1993).
[3] L. Gao, Y. Y. Xue, F. Chen, Q. Xiong, R. L. Meng, D. Ramirez, C. W. Chu, J. H. Eggert, and H. K. Mao, Phys. Rev. B(R) **50**, 4260 (1994).
[4] M. Rotter, M. Tegel, and D. Johrendt, Phys. Rev. Lett. **101**, 107006 (2008).

[5]  K. Sasmal, B. Lv, B. Lorenz, A. M. Guloy, F. Chen, Y. Y. Xue, and C. W. Chu, Phys. Rev. Lett. **101**, 107007 (2008).

[6]  J. H. Tapp, Z. Tang, B. Lv, K. Sasmal, B. Lorenz, P. C. W. Chu, and A. M. Guloy, Phys. Rev. B(R) **78**, 060505 (2008).

[7]  F. C. Hsu, J. Y. Luo, K. W. Yeh, T. K. Chen, T. W. Huang, P. M. Wu, Y. C. Lee, Y. L. Huang, Y. Y. Chu, D. C. Yan, and M. K. Wu, Proc. Natl. Acad. Sci. U.S.A. **105**, 14262 (2008).

[8]  K. Zhao, Q. Q. Liu, X. C. Wang, Z. Deng, Y. X. Lv, J. L. Zhu, F. Y. Li, and C. Q. Jin, Journal of physics. Condensed matter : an Institute of Physics journal **22**, 222203 (2010).

[9]  P. L. Alireza, Y. T. Ko, J. Gillett, C. M. Petrone, J. M. Cole, G. G. Lonzarich, and S. E. Sebastian, Journal of physics. Condensed matter : an Institute of Physics journal **21**, 012208 (2009).

[10] Y. Zheng, Y. Wang, B. Lv, C. W. Chu, and R. Lortz, New Journal of Physics **14**, 053034 (2012).

[11] W. Yu, A. A. Aczel, T. J. Williams, S. L. Bud'ko, N. Ni, P. C. Canfield, and G. M. Luke, Phys. Rev. B(R) **79**, 020511 (2009).

[12] A. Kreyssig, M. A. Green, Y. Lee, G. D. Samolyuk, P. Zajdel, J. W. Lynn, S. L. Bud'ko, M. S. Torikachvili, N. Ni, S. Nandi, J. B. Leão, S. J. Poulton, D. N. Argyriou, B. N. Harmon, R. J. McQueeney, P. C. Canfield, and A. I. Goldman, Phys. Rev. B **78**, 184517 (2008).

[13] B. Lv, L. Z. Deng, M. Gooch, F. Wei, Y. Sun, J. K. Meen, Y. Y. Xue, B. Lorenz, and C. W. Chu, Proc. Natl. Acad. Sci. U.S.A. **108**, 15705 (2011).

[14] S. R. Saha, N. P. Butch, T. Drye, J. Magill, S. Ziemak, K. Kirshenbaum, P. Y. Zavalij, J. W. Lynn, and J. Paglione, Phys. Rev. B **85**, 024525 (2012).

[15] Z. Gao, Y. Qi, L. Wang, D. Wang, X. Zhang, C. Yao, C. Wang, and Y. Ma, EPL (Europhysics Letters) **95**, 67002 (2011).

[16] Y. Qi, Z. Gao, L. Wang, D. Wang, X. Zhang, C. Yao, C. Wang, C. Wang, and Y. Ma, Supercon. Sci. Technol. **25**, 045007 (2012).

[17] F. Wei, B. Lv, L. Deng, J. K. Meen, Y. Y. Xue, and C. W. Chu, Philos. Mag. **94**, 2562 (2014).

[18] K. Gofryk, M. Pan, C. Cantoni, B. Saparov, J. E. Mitchell, and A. S. Sefat, Phys. Rev. Lett. **112**, 047005 (2014).

[19] I. Zeljkovic, D. Huang, C. L. Song, B. Lv, C. W. Chu, and J. E. Hoffman, Phys. Rev. B(R) **87**, 201108 (2013).

[20] W. Uhoya, D. Cargill, K. Gofryk, G. M. Tsoi, Y. K. Vohra, A. S. Sefat, and S. T. Weir, High Pressure Research **34**, 49 (2014).

[21] S. R. Saha, T. Drye, S. K. Goh, L. E. Klintberg, J. M. Silver, F. M. Grosche, M. Sutherland, T. J. S. Munsie, G. M. Luke, D. K. Pratt, J. W. Lynn, and J. Paglione, Phys. Rev. B **89**, 134516 (2014).

[22] K. W. Lee, V. Pardo, and W. E. Pickett, Phys. Rev. B **78**, 174502 (2008).

[23] Panayiotis A. Varotsos and K. D. Alexopoulos, *Thermodynamics of point defects and their relation with bulk properties* (North-Holland, 1986).

[24] W. Li, H. Ding, P. Deng, K. Chang, C. Song, K. He, L. Wang, X. Ma, J.-P. Hu, X. Chen, and Q.-K. Xue, Nature Physics **8**, 126 (2011).

[25] A. Charnukha, A. Cvitkovic, T. Prokscha, D. Propper, N. Ocelic, A. Suter, Z. Salman, E. Morenzoni, J. Deisenhofer, V. Tsurkan, A. Loidl, B. Keimer, and A. V. Boris, Phys. Rev. Lett. **109**, 017003 (2012).

[26] A. F. May, M. A. McGuire, H. Cao, I. Sergueev, C. Cantoni, B. C. Chakoumakos, D. S. Parker, and B. C. Sales, Phys. Rev. Lett. **109**, 077003 (2012).

[27] R. B. Goldfarb and C. E. Patton, Phys. Rev. B **24**, 1360 (1981).

[28] N. Paunovic, Z. V. Popovic, and Z. D. Dohcevic-Mitrovic, J. Phys. Condens. Matter **24**, 456001 (2012).

[29] T. Okada, H. Ogino, H. Yakita, A. Yamamoto, K. Kishio, and J. Shimoyama, Physica C: Superconductivity **505**, 1 (2014).

[30] C. S. Lue, J. H. Ross, C. F. Chang, and H. D. Yang, Phys. Rev. B(R) **60**, 13941 (1999).

[31] C. W. Chu, B. Lv, L. Z. Deng, B. Lorenz, B. Jawdat, M. Gooch, K. Shrestha, K. Zhao, X. Y. Zhu, Y. Y. Xue, and F. Y. Wei, J. Phys.: Conf. Ser. **449**, 012014 (2013).
[32] B. Lv, L. Deng, Z. Wu, F. Wei, K. Zhao, J. K. Meen, Y. Y. Xue, L. L. Wang, X. C. Ma, Q. K. Xue, and C. W. Chu, MRS Proc. **1684**, mrss14-1684-t04-01 doi:10.1557/opl.2014.880 (2014).
[33] P. Zavalij, W. Bao, X. F. Wang, J. J. Ying, X. H. Chen, D. M. Wang, J. B. He, X. Q. Wang, G. F. Chen, P. Y. Hsieh, Q. Huang, and M. A. Green, Phys. Rev. B **83**, 132509 (2011).